 \documentclass[twocolumn]{webofc}
\usepackage{natbib}
\usepackage[varg]{txfonts}   
\usepackage{hyperref}
\usepackage{url}

\hypersetup{colorlinks=true,citecolor=blue,urlcolor=blue,linkcolor=blue}

\begin{document}

\title{Test-beam measurements of instrumented sensor planes for a highly compact and granular electromagnetic calorimeter}

\author{\firstname{Melissa} \lastname{Almanza Soto}\inst{1}\fnsep\thanks{\email{melissa.almanza@ific.uv.es}} on behalf of the LUXE Collaboration
}

\institute{IFIC, CSIC and Universitat de Val\`encia, C/ Catedr\`atic Jos\'e Beltr\'an Mart\'inez 2,
46980 Paterna, Spain
}

\abstract{

The LUXE experiment is designed to explore the strong-field QED regime in interactions of high-energy electrons from the European XFEL in a powerful laser field. One of the crucial aims of this experiment is to measure the production of electron-positron pairs as a function of the laser field strength where non-perturbative effects are expected to kick in above the Schwinger limit.
For the measurements of positron energy and multiplicity spectra, a tracker and an electromagnetic calorimeter are foreseen. 
The expected number of positrons varies over ten orders of magnitude and has to be measured over a widely spread low-energy background. To overcome these challenges, a compact and finely segmented calorimeter is proposed. The concept of a sandwich calorimeter made of tungsten absorber plates interspersed with thin sensor planes is developed.
The sensor planes comprise a silicon pad sensor, flexible Kapton printed circuit planes for bias voltage supply and signal transport to the sensor edge, all embedded in a carbon fiber support. The thickness of a sensor plane is less than 1 mm. A dedicated readout is developed comprising front-end ASICs in 130 nm technology and FPGAs to orchestrate the ASICs and perform data pre-processing. As an alternative, GaAs are considered with integrated readout strips on the sensor. Prototypes of both sensor planes are studied in an electron beam of 5 GeV. Results will be presented on the homogeneity of the response and edge effects.

}
\maketitle
\section{Introduction}
\label{intro}
The Laser Und XFEL Experiment (LUXE) is an experiment proposal with the mission of studying the behavior of Quantum Electrodynamics (QED) in the strong-field regime. In particular, LUXE will take precision measurements of electron-photon and photon-photon interactions in the transition from perturbative to non-perturbative QED. LUXE also opens the possibility to perform a sensitive search for new particles beyond the Standard Model that couple to photons \cite{CDR2021}. 

The experiment will use the 16.5 GeV electron beam from the European X-Ray Free-Electron Laser and collide it with a high-intensity laser to reach and surpass the critical field strength $\mathcal{E}_{\text{cr}}={m_e}^2c^3/(e\hbar)=1.32\times 10^{18}\ \text{V/m}$, known as the Schwinger limit. 

In its initial phase, the experiment will use an existing 40~TW laser. Later on, the laser power will be upgraded to 350~TW for the second phase. LUXE will use the 10~Hz electron bunch rate of XFEL, with collisions occurring only at 1 Hz to match the laser pulse rate. The remaining electron bunches will be used for background measurements \cite{CDR2021}.   

The experiment is planned to have two modes which involve the production of electrons, positrons, and photons. In the first mode, beam electrons will collide directly with the high-power tightly-focused laser pulse. For the second mode, the high energy electrons will first be converted into a photon beam and then this photons will interact with the laser pulse.
The multiplicities and energies of the particles produced during the interactions will be measured by a set of detectors. An electromagnetic calorimeter and a tracker compose the system to measure the number of positrons and their energy spectrum. 

\section{The Positron Electromagnetic Calorimeter for LUXE}
\label{ecalLUXE}
For both modes of the experiment, the same electromagnetic calorimeter is foreseen for the positron side of LUXE, this detector is known as ECAL-P. 

The main challenge in the design of ECAL-P is the vast range of multiplicities for positrons between the two modes of the experiment. The expected positron rates vary from $10^{-4}$ to more than $10^6$ per bunch crossing \cite{TDRLUXE2023}.
In the low multiplicities expected for the $\gamma-$laser mode, the calorimeter must be able to perform measurements with an excellent background rejection. In contrast, for the high positron multiplicities of the e-laser mode the detector must achieve a good linearity. 

The proposed solution consists of a compact sampling calorimeter with a small Moli\`ere radius and a high granularity. 
The ECAL-P design consists of tungsten absorber plates and thin active layers.  Tungsten was chosen as the absorber material, as it has a small Moli\`ere radius of 9.3~mm. The detector will have 21 tungsten absorber layers with a thickness of 3.5~mm, equivalent to one radiation length. The active layers are to be kept to less than 1~mm in thickness.

Two sensor materials are proposed for the sensitive layers: Gallium Arsenide (GaAs) and Silicon.

The GaAs sensors were produced by the National Tomsk State University. The sensor is made out of a single GaAs crystal compensated with chromium. Each pad has an area of $4.7 \times 4.7 ~\text{mm}^2$ with a separation of 0.3~mm between pads. A distinctive feature of the GaAs sensors is the presence of aluminum traces embedded in the gaps between the pads. The aluminum traces are connected to the pads and take the signal to the top edge of the sensor plane. The metal traces favor the compact design of the calorimeter by keeping all the electronics on top of the detector and outside of the absorber and the active volume of ECAL-P. 
An advantage of GaAs over silicon is that the former has a higher radiation tolerance, this being the reason why it has been proposed for other calorimeters such as BeamCal for the ILD \cite{forward}, detector in which the ECAL-P design has been based on.

\begin{figure}[h]
\centering
\includegraphics[width=5cm,clip]{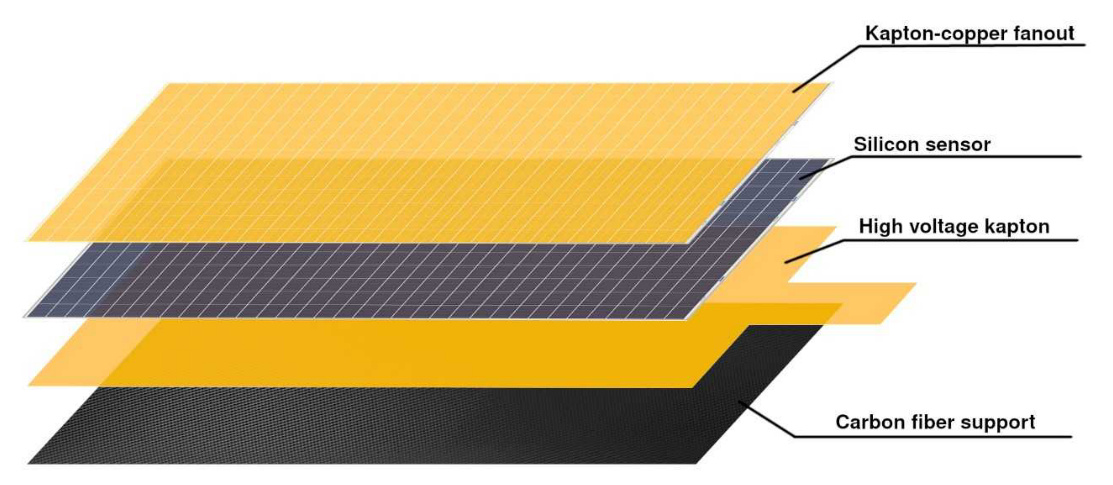}
\caption{Sketch of the components of an instrumented sensor plane for the LUXE ECAL-P.}
\label{siliconLayers}       
\end{figure}

The silicon sensors are produced by Hamamatsu. The sensor pads have a $5.5 \times 5.5 ~\text{mm}^2$ area and a $320~\mu \text{m}$ thickness. The space between pads is 0.01~mm. A flexible Kapton PBC with copper pads and traces connected to the sensor pads with conductive glue will guide the signal to the edge of the sensor. The sub-layer components of a silicon sensor instrumented plane are shown in Fig.~\ref{siliconLayers}. Similar signal transport techniques have been tested in various geometries of the LumiCAL detector prototypes \cite{LumiCAL2019},\cite{moliere2018}. 

The front-end ASICs for the readout of the sensor planes of ECAL-P are called Fcal ASIC for XFEL Experiment (FLAXE). These ASICs are based on the FCAL ASIC for Multiplane Readout (FLAME), designed for the silicon sensors of the LumiCAL detector. The FLAME ASICs have been used in several beam-tests of the FCAL collaboration and were used by the LUXE ECAL-P group in the beam-test described in the following section.

\section{Test-beam measurements of active planes}\label{testbeam}
A beam-test of four prototypes of the ECAL-P sensor planes was performed at the DESY-II test-beam facility in September of 2022. Two $16~\times~8$ pad arrays of silicon sensors and two $15~\times~10$ pad arrays of GaAs sensor planes were put to test in a 5~GeV electron beam. The silicon sensors tested were 500~$\mu $m thick, differing from the $320~\mu $m thick sensors proposed for the ECAL-P. 

The FLAME ASICs were used for the readout of the sensor planes in the beam-test. 
These ASICs are 32-channel ASICs in CMOS 130~nm with analog front-end and 10-bit ADC in each channel, followed by two fast (5.2~Gbps) serializers and data transmitters, more specifications for each component can be found in table~\ref{tab:asic}.

\begin{table}
\centering
\caption{Specifications for FLAME front-end ASIC}
\label{tab:asic}       
\begin{tabular}{ll}
\hline
\multicolumn{1}{c}{Element} & \multicolumn{1}{c}{Specifications}          \\ \hline
Analog front-end in channel & $\bullet$CR-RC shaping $T_{peak}\sim 50$ns  \\
                            & $\bullet$Switched gain \\
                            & $\bullet C_{in}$ 20-40pF                    \\
                            &                                             \\
10-but ADC in channel       & $\bullet f_{sample}=20$MHz                  \\
                            & $\bullet$ENOB > 9.5                         \\ \hline
\end{tabular}
\end{table}

A beam telescope composed of 6 Alpide sensors was used to measure the tracks of the beam electrons and predict their entry point on the sensor plane. All 6 telescope planes were set in between a beam collimator and the sensor plane under test.  

\subsection{Performance of the sensor planes}\label{performance}
The readout channels used for the beam-test were calibrated for differences in the pre-amplification, these calibration factors were applied to the test-beam data. 

\begin{figure}[ht]
\centering
\sidecaption
\includegraphics[trim={1cm 0.4cm 0 0},width=5cm,clip]{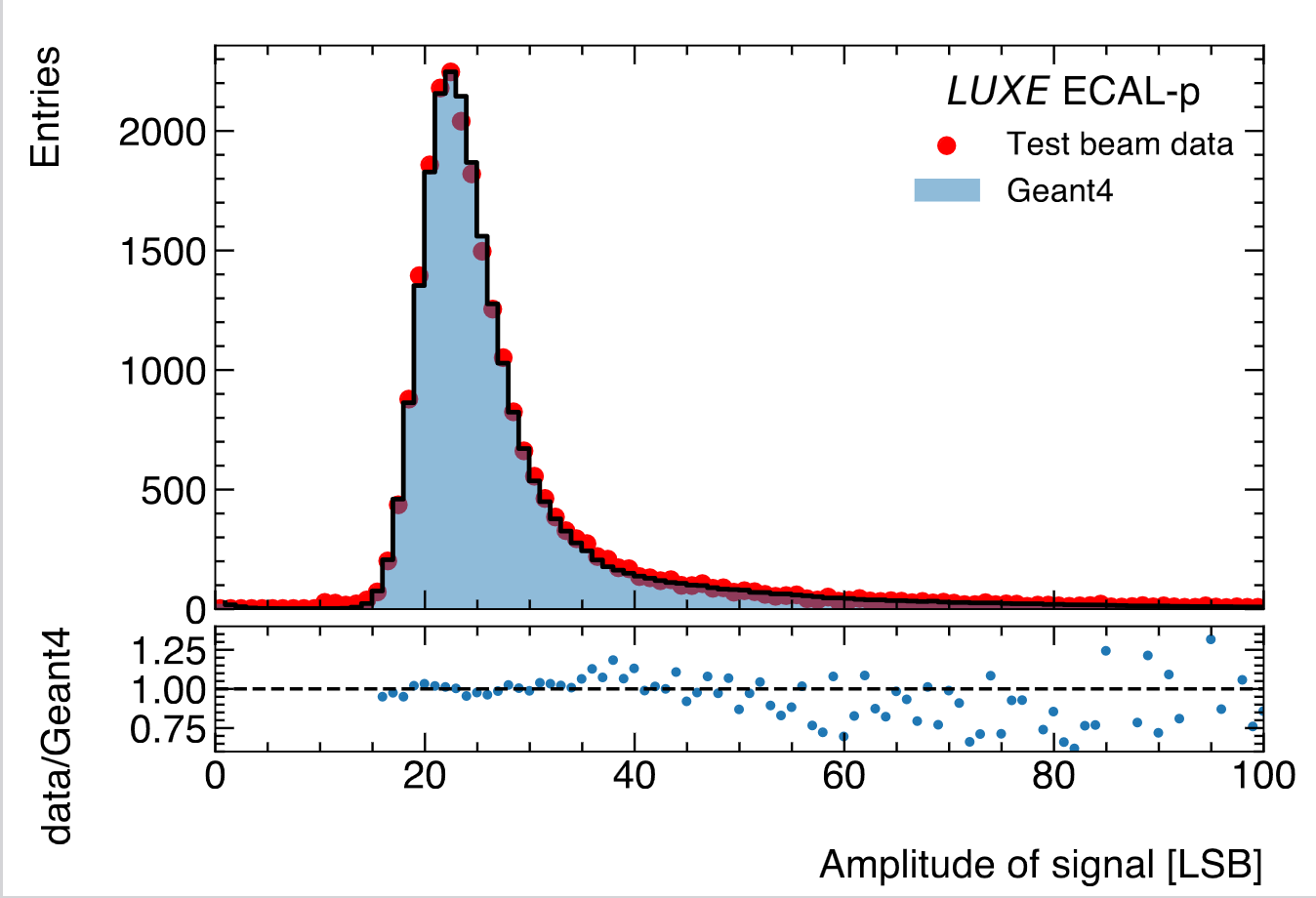}
\caption{Signal distribution of 5~GeV electrons in the silicon sensor. Comparison between test-beam data and simulation.}
\label{montecarlo} 
\end{figure}

The response of the silicon sensor plane to 5~GeV electrons was simulated using Geant4. The simulation included the full material of the sensor planes (carbon fiber frame, flex PCB, sensor, high-voltage flex PCB). As shown in Fig.~\ref{montecarlo}, the simulation shows a good agreement with the test-beam measurements around the Most Probable Value (MPV). 

\begin{figure}[ht]
\centering
\sidecaption
\includegraphics[width=5cm,clip]{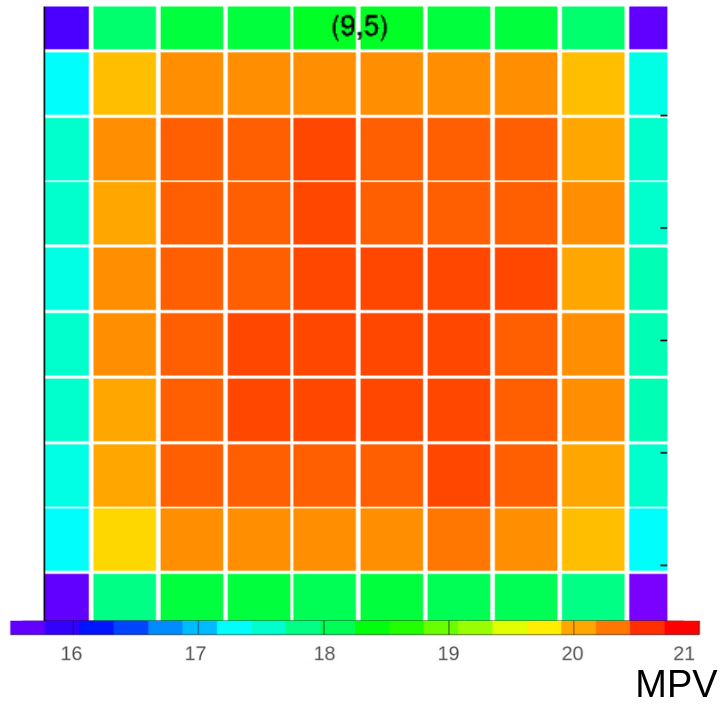}
\caption{Map of the MPV in a pad of the GaAs sensor.}
\label{mpvgaas2d} 
\end{figure}

\begin{figure}[ht]
\centering
\sidecaption
\includegraphics[width=5cm,clip]{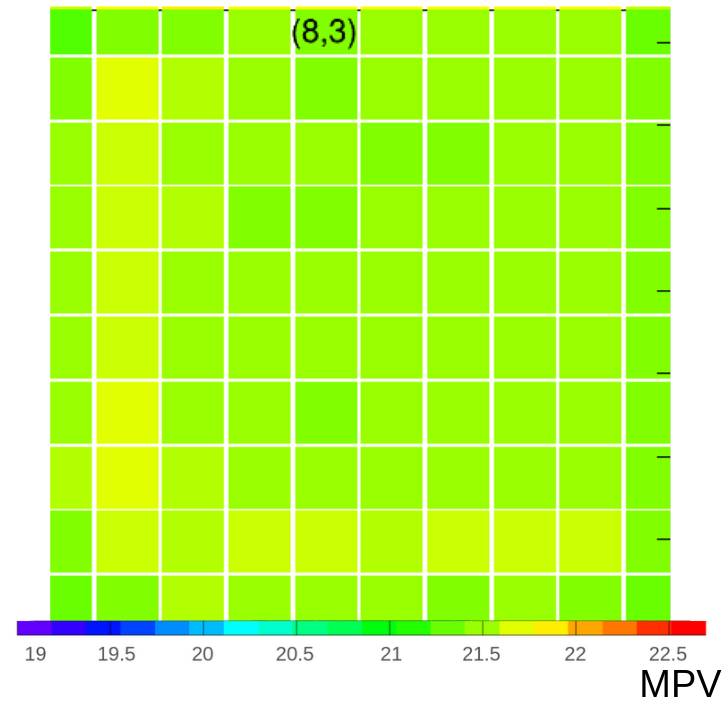}
\caption{Map of the MPV in a pad of the silicon sensor.}
\label{mpvsi2d} 
\end{figure}

Using the telescope data, the area of each pad was subdivided into pixels. The signal distribution of electrons hitting each section was plotted and the MPV was calculated using a Landau function convoluted with a Gaussian fit. 
The MPV distribution as a function of the pad's local X and Y coordinates shows a drop in the MPV when approaching the edges of the GaAs sensor pad, as seen in Fig.~\ref{mpvgaas2d}. As for the silicon sensors, Fig.~\ref{mpvsi2d} shows an L-shaped area with a $1\%$ rise in amplitude with respect to the average. The same effect was observed for all the pads studied.

\begin{figure}[ht]
\centering
\sidecaption
\includegraphics[width=5cm,clip]{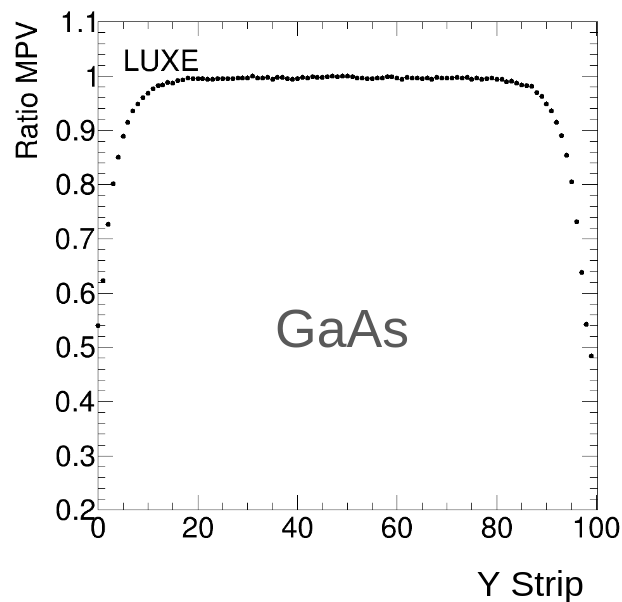}
\caption{The MPV in a pad normalized to the MPV of the center of the pad as a function of the local vertical coordinate for the GaAs sensor.}
\label{GAASratio} 
\end{figure}

\begin{figure}[ht]
\centering
\sidecaption
\includegraphics[width=5cm,clip]{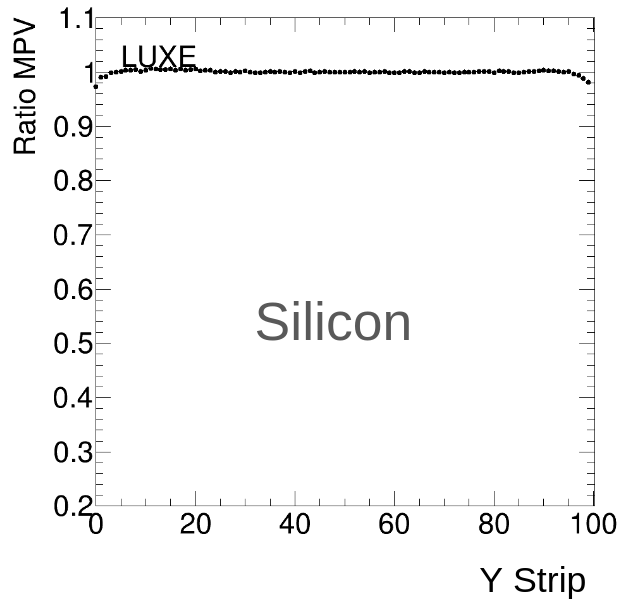}
\caption{The MPV in a pad normalized to the MPV of the center of the pad as a function of the local vertical coordinate for the silicon sensor.}
\label{siratio} 
\end{figure}

To quantify the edge effects, the pad area was separated into thin strips. The MPV of the signal distribution of electrons hitting each strip area was calculated. At the very edge of the GaAs pad, the drop amounts to $50\%$ of the MPV of the center of the pad. The drop in signal is shown for a typical pad in Fig.~\ref{GAASratio}. A drop of 2 to 3$\%$ at the edges of the pad can be observed in Fig.~\ref{siratio} for the silicon sensor.

\begin{figure}[ht]
\centering
\sidecaption
\includegraphics[width=5cm,clip]{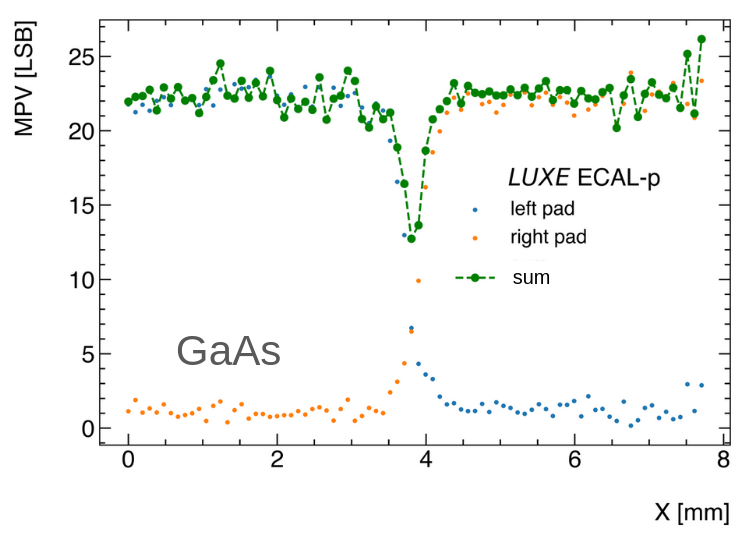}
\caption{Signal of electrons in two neighboring pads as a function of the local horizontal coordinate for the GaAs sensor.}
\label{gaasdropx}       
\end{figure}

A study was performed in which the signals from two adjacent pads were added for events in which electrons are predicted to enter the detector along the gap between the two pads. Scanning along the horizontal coordinate for the GaAs sensor, one observes a maximum drop in signal of $42\%$ with respect to the average, as shown in Fig.~\ref{gaasdropx}.

Along the vertical coordinate, the signal decreases $13\%$ with respect to the average. This effect was not observed for the silicon sensor along either coordinate. 

\begin{figure}[ht]
\centering
\sidecaption
\includegraphics[width=5cm,clip]{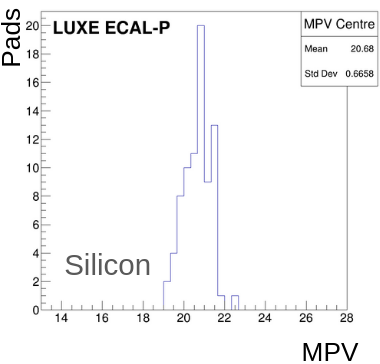}
\caption{Distribution of the MPV for electrons hitting the center of pads for a silicon sensor.}
\label{homSILICON}       
\end{figure}

\begin{figure}[ht]
\centering
\sidecaption
\includegraphics[width=5cm,clip]{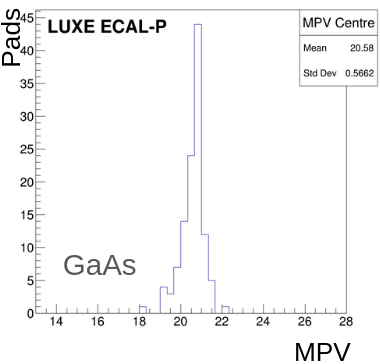}
\caption{Distribution of the MPV for electrons hitting the center of pads for a GaAs sensor.}
\label{homGAAS}       
\end{figure}

Lastly, the MPV of the signal distribution was calculated for every pad in a sensor plane. To discard the edge effects of the GaAs pads and make a comparison between the two types of sensors, only the signals generated from electrons predicted to hit the center area of a pad were used. The distributions of the MPV are shown in Figs.~\ref{homSILICON} and \ref{homGAAS} for the silicon and the GaAs sensor, respectively.
The results show that the GaAs sensor has an overall larger MPV than that of silicon, this is due to the higher density of the material. 
 
The homogeneity of the response of the GaAs sensor was found to be comparable to that of silicon after discarding electrons hitting the edges of pads. The homogeneity of the response in all pads is within 3\% for both sensors. 

\section{Conclusion}\label{conclusion}
 Individual pad-response studies were possible using the beam telescope. The GaAs sensors with the aluminium traces present edge-effects involving a drop in the amplitude of signals. The same effect is found for the silicon sensor in a smaller scale. Gain calibration factors were applied to the test-beam data. After removing the edges from the GaAs sensor pads, the homogeneity of the response was found to be comparable to that of the silicon sensor. 
A Monte Carlo simulation of the instrumented silicon plane was performed and was found to be a good modeling of the test-beam data.

\section{Acknowledgements}
We thank the DESY technical staff for continuous assistance and the DESY directorate for their strong support and the hospitality they extend to the non-DESY members of the collaboration. This work has benefited from computing services provided by the German National Analysis Facility (NAF) and the Swedish National Infrastructure for Computing (SNIC). The measurements leading to these results have been performed at the Test Beam Facility at DESY Hamburg (Germany), a member of the Helmholtz Association (HGF). M.A.S. acknowledges the financial support by Spanish MICIU/ AEI and European Union / FEDER via the grant PID2021-122134NB-C21. EM and by Generalitat Valenciana (GV) via the Excellence Grant CIPROM/2021/073. M.A.S also acknowledges the support from the MCIN with funding from the European Union NextGenerationEU and Generalitat Valenciana in the call Programa de Planes Complementarios de I+D+i (PRTR 2022) through the project with reference ASFAE/2022/015.
\bibliography{bibliography}

\begin{thebibliography}{5}

\bibitem{CDR2021}
{H. Abramowicz et al.}, Conceptual design report for the luxe experiment, Eur. Phys. J. Spec. Top. \textbf{230}, 2445 (2021). \doiwoc{10.1140/epjs/s11734-021-00249-z}

\bibitem{TDRLUXE2023}
{H. Abramowicz et al.}, Technical design report for the {LUXE} experiment (2023). \doiwoc{10.48550/arXiv.2308.00515}

\bibitem{forward}
{H. Abramowicz et al.}, Forward instrumentation for {ILC} detectors, JINST \textbf{5}, P12002 (2010). \doiwoc{10.1088/1748-0221/5/12/P12002}

\bibitem{LumiCAL2019}
{H. Abramowicz et al.}, Performance and {Molière} radius measurements using a compact prototype of {LumiCal} in an electron test beam, European Physical Journal C \textbf{79}, 579 (2019). \doiwoc{10.1140/epjc/s10052-019-7077-9}

\bibitem{moliere2018}
{H. Abramowicz et al.}, Measurement of shower development and its {Molière} radius with a four-plane {LumiCal} test set-up, Eur. Phys. J. C \textbf{78}, 135 (2018). \doiwoc{10.1140/epjc/s10052-018-5611-9}

\end{thebibliography}
\end{document}